\documentclass[runningheads]{llncs}
\usepackage{graphicx}
\usepackage{amsmath}
\usepackage{amssymb}
\usepackage{tikz-cd}
\usepackage{amsfonts}
\usepackage[ruled,vlined]{algorithm2e}
\usepackage{wrapfig}
\usepackage{url}
\usepackage{multirow}
\usepackage{booktabs} 
\usepackage{bbding}

\begin{document}
\title{Cortical Morphometry Analysis based on Worst Transportation Theory}
\titlerunning{Cortical Morphometry Analysis based on Worst Transportation Theory}
%
\author{
Min Zhang\inst{1\star} \and
Dongsheng An\inst{2\star}  \and
Na Lei\inst{3}\textsuperscript{(\Envelope)} \and
Jianfeng Wu \inst{4} \and
Tong Zhao \inst{5} \and
Xiaoyin Xu \inst{6} \and
Yalin Wang \inst{4} \and
Xianfeng Gu \inst{2}
}

\authorrunning{Zhang et al.}
%
\institute{
Zhejiang University\\
\email{min\_zhang@zju.edu.cn} \and
Department of Computer Science, Stony Brook University, Brookhaven, USA\\
\email{\{doan,gu\}@cs.stonybrook.edu}\and
DUT-RU ISE, Dalian University of Technology \\
\email{nalei@dlut.edu.cn} \and
School of Computing, Informatics, and Decision Systems Engineering,
Arizona State University, Arizona, USA \\
\email{\{jianfen6,ylwang\}@asu.edu}\and
Université Côte d'Azur, Inria \\
\email{tong.zhao@inria.fr}\and
Brigham and Women's Hospital, Harvard Medical School\\
\email{xxu@bwh.harvard.edu}
}

\renewcommand{\thefootnote}{\fnsymbol{footnote}}
\footnotetext[1]{The two authors contributed  equally to this paper.}
\maketitle              
\begin{abstract} 
Biomarkers play an important role in early detection and intervention in Alzheimer’s disease (AD). However, obtaining effective biomarkers for AD is still a big challenge. In this work, 
we propose to use the worst transportation cost as a univariate biomarker to index cortical morphometry for tracking AD progression. 
The worst transportation (WT) aims to find the least economical way to transport one measure to the other, which contrasts to the optimal transportation (OT) that finds the most economical way between measures. 
To compute the WT cost, we generalize the Brenier theorem for the OT map to the WT map, and show that the WT map is the gradient of a concave function satisfying the Monge-Ampere equation. We also develop an efficient algorithm to compute the WT map based on computational geometry. We apply the algorithm to analyze cortical shape difference between dementia due to AD and normal aging individuals. The experimental results reveal the effectiveness of our proposed method which yields better statistical performance than other competiting methods including the OT. 


\end{abstract}
\section{Introduction}
As the population living longer, Alzheimer’s disease (AD) is now a major public health concern with the number of patients expected to reach 13.8 million by the year 2050 in the U.S. alone \cite{Brookmeyer2007ADforcast}. However, the late interventions or the targets with secondary effects and less relevant to the disease initiation often make the current therapeutic failures in patients with dementia due to AD \cite{Hyman2011Amyloid}. 
The accumulation of beta-amyloid plaques (A$\beta$) in human brains is one of the hallmarks of AD, and preclinical AD is now viewed as a gradual process before the onset of the clinical symptoms. The A$\beta$ positivity is treated as the precursor of anatomical abnormalities such as atrophy and functional changes such as hypometabolism/hypoperfusion. 

It is generally agreed that accurate presymptomatic diagnosis and preventative treatment of AD could have enormous public health benefits. Brain A$\beta$ pathology can be measured using positron emission tomography (PET) with amyloid-sensitive radiotracers, or in cerebrospinal fluid (CSF). However, these invasive and expensive measurements are less attractive to subjects in preclinical stage and PET scanning is also not widely available in clinics. Therefore, there is strong interest to develop structural magnetic resonance imaging (MRI) biomarkers, which are largely accessible, cost-effective and widely used in AD clinical research, to predict brain amyloid burden~\cite{Ansart:amyloidosis}. Tosun et al.~\cite{Tosun2014MRI} combine MRI-based measures of cortical shape and cerebral blood flow to predict amyloid status for early-MCI individuals. Pekkala et al.~\cite{Pekkala2020Amyloid} use the brain MRI measures like volumes  of  the cortical  gray  matter,  hippocampus,  accumbens,  thalamus  and  putamen to identify the A$\beta$ positivity in cognitively unimpaired (CU) subjects.
Meanwhile, a univariate imaging biomarker would be highly desirable for clinical use and randomized clinical trials (RCT)\cite{Shi:TPAMI20,Tu:NEIN20}. Though a variety of research studies univariate biomarkers with sMRI analysis, there is limited research to develop univariate biomarker to predict brain amyloid burden, which will enrich our understanding of the relationship between brain atrophy and AD pathology and thus benefit assessing disease burden, progression and effects of treatments.

In this paper, we propose to use the worst transportation (WT) cost as a univariate biomarker to predict brain amyloid burden. Specifically, we compare the population statistics of WT costs
between the A$\beta+$ AD and the A$\beta-$ CU subjects. The new proposed WT transports one measure to the other in the least economical way, which contrasts to the optimal transportation (OT) map that finds the most economical way to transport from one measure to the other. Similar to the OT map, the WT map is the gradient of a strictly concave function, which also satisfies the Monge-Ampere equation. Furthermore, the WT map can be computed by convex optimization with the geometric variational approach like the OT map \cite{gu2013theory,Su2015Registration}. 
Intuitively, the OT/WT maps are solely determined by the Riemannian metric of the cortical surface, and they reflect the intrinsic geometric properties of the brain. Therefore, they tend to serve as continuous and refined shape difference measures.

\paragraph{Contributions} The contribution of the paper includes:
(i) In this work, we generalize the Brenier theorem from the OT map to the newly proposed WT map, and rigorously show that the WT map is the gradient of a concave function that satisfies the Monge-Ampere equation. To the best of our knowledge, it is the first WT work in medical imaging research; (ii) We propose an efficient and robust computational algorithm to compute the WT map based on computational geometry. We further validate it with geometrically complicated human cerebral cortical surfaces;
(iii) Our extensive experimental results show that the proposed WT cost performs better than the OT cost when discriminating A$\beta+$ AD patients from A$\beta-$ CU subjects. This surprising result may help broaden univariate imaging biomarker research by opening up and addressing a new theme.

\vspace{-5mm}
\section{Theoretic Results}
\vspace{-3mm}
\label{sec:theory}
In this section, we briefly review the theoretical foundation of optimal transportation, then generalize the Brenier theorem 
and Yau's theorem to the worst transportation.
\vspace{-4mm}
\subsection{Optimal Transportation Map}
\vspace{-1mm}
Suppose $\Omega, \Omega^*\subset \mathbb{R}^d$  are domains in Euclidean space, with probability measures $\mu$ and $\nu$ respectively satisfying the equal total mass condition: $\mu(\Omega) = \nu(\Omega^*)$. The density functions are $d\mu=f(x)dx$ and $d\nu=g(y)dy$. The transportation map $T:\Omega\to\Omega^*$ is \emph{measure preserving} if for any Borel set $B\subset \Omega^*$, $\int_{T^{-1}(B)} d\mu(x) = \int_B d\nu(y)$, denoted as $T_\#\mu=\nu$. \\
Monge raised the \emph{optimal transportation map problem} \cite{villani2008optimal}: given a \emph{transportation cost function} $c:\Omega\times \Omega^*\to \mathbb{R}^+$, find a transportation map $T:\Omega\to\Omega^*$ that minimizes the \emph{total transportation cost},
\[
(MP)\quad    \min_T\left\{ \int_\Omega c(x,T(x)): T:\Omega\to\Omega^*, T_\#\mu = \nu \right\}.
\]
The minimizer is called the \emph{optimal transportation map} (OT map). The total transportation cost of the OT map is called the \emph{OT cost}.

\begin{theorem}[Brenier~\cite{Brenier_1991}]
Given the measures $\mu$ and $\nu$ with compact supports $\Omega,\Omega^*\subset \mathbb{R}^d$ with equal total mass $\mu(\Omega)=\nu(\Omega^*)$, the corresponding density functions $f,g\in L^1(\mathbb{R}^d)$, and the cost function $c(x,y)=\frac{1}{2}|x-y|^2$, then the optimal transportation map $T$ from $\mu$ to $\nu$ exists and is unique. It is the gradient of a convex function $u:\Omega\to\mathbb{R}$, the so-called Brenier potential. $u$ is unique up to adding a constant, and $T=\nabla u$.
\label{thm:Brenier}
\end{theorem}
If the Brenier potential is $C^2$, then by the measure preserving condition, it satisfies the Monge-Amp\`ere equation,
\begin{equation}
    \text{det} D^2u(x) = \frac{f(x)}{g\circ \nabla u(x)}.
    \label{eqn:Monge_ampere}
\end{equation}
where $D^2u$ is the Hessian matrix of $u$.
\vspace{-4mm}
\subsection{Worst Transportation Map}
\vspace{-1mm}
With the same setup, the worst transportation problem can be formulated as follows: given the transportation cost function $c:\Omega\times \Omega^*\to \mathbb{R}^+$, find a measure preserving map $T:\Omega\to\Omega^*$ that maximizes the total transportation cost,
\[
(WP)\quad    \max_T\left\{ \int_\Omega c(x,T(x)): T:\Omega\to\Omega^*, T_\#\mu = \nu \right\}.
\]
The maximizer is called the \emph{worst transportation map}. The transportation cost of the WT map is called the \emph{worst transportation cost} between the measures. In the following, we generalize the Brenier theorem to the WT map.

\begin{theorem}[Worst Transportation Map]
Given the probability measures $\mu$ and $\nu$ with compact supports $\Omega,\Omega^*\subset \mathbb{R}^{d}$ respectively with equal total mass $\mu(\Omega)=\nu(\Omega^*)$, and assume the corresponding density functions $f,g\in L^1(\mathbb{R}^d)$, the cost function $c(x,y)=\frac{1}{2}|x-y|^2$, then the {\bf worst} transportation map exists and is unique. It is the gradient of a {\bf concave} function $u:\Omega\to\mathbb{R}$, where $u$ is the worst Brenier potential function, unique up to adding a constant. The WT map is given by $T=\nabla u$. Furthermore, if $u$ is $C^2$, then it satisfies the Monge-Amp\`ere equation in Eqn.~(\ref{eqn:Monge_ampere}).
\label{thm:Brenier_wt}
\end{theorem}

\begin{proof} Suppose $T: \Omega \to \Omega^*$ is a measure-preserving map, $T_\#\mu=\nu$. Consider the total transportation cost, 
{\footnotesize
\[
\begin{split}
&\int_\Omega |x-T(x)|^2 d\mu = \int_\Omega |x|^2d\mu + \int_\Omega |T(x)|^2 d\mu - 2\int_\Omega\langle x,T(x)\rangle d\mu\\
&=\int_\Omega |x|^2d\mu + \int_{\Omega^*} |y|^2 d\nu - 2\int_\Omega\langle x,T(x)\rangle d\mu, \quad \text{with}~y=T(x).
\end{split}
\]
}
Therefore, maximizing the transportation cost is equivalent to 
$\min_{T_\#\mu=\nu} \int_\Omega \langle x, T(x) \rangle d\mu$. With the Kantorovich formula, this is equivalent to finding the following transportation plan $\gamma:\Omega\times\Omega^*\to\mathbb{R}$, 
\[
\min_\gamma\left\{\int_{\Omega\times\Omega^*} \langle x,y \rangle d\gamma,
(\pi_x)_\#\gamma=\mu,
(\pi_y)_\#\gamma=\nu
\right\},
\]
where $\pi_x$, $\pi_y$ are the projections from $\Omega\times\Omega^*$ to $\Omega$ and $\Omega^*$ respectively. By duality, this is equivalent to $\max\{ J(u,v), (u,v)\in K\}$, where the energy $J(u,v):= \int_\Omega u(x)f(x)dx + \int_{\Omega^*} v(y) g(y)dy$, and the functional space $K:=\left\{(u,v):u(x)+v(y)\le \langle x, y \rangle \right\}$. Now we define the c-transform,
\begin{equation}
    u^c(y):= \inf_{x\in \bar{\Omega}} \langle x,y \rangle - u(x).
    \label{eqn:c_transform}
\end{equation}
Fixing $x$, $\langle x, y \rangle - u(x)$ is a linear function, hence $u^c(y)$ is the lower envelope of a group of linear functions, and thus is a concave function with Lipschitz condition (since the gradient of each linear function is $x\in \bar{\Omega}$, $\bar{\Omega}$ is bounded). We construct a sequence of function pairs $\{(u_k,v_k)\}$, where $u_k = v_{k-1}^c$, $v_k=u_k^c$. Then $J(u_k,v_k)$ increases monotonously, and the Lipschitz function pairs $(u_k,v_k)$ converge to $(u,v)$, which is the maximizer of $J$. Since $u$ and $v$ are c-transforms of each other, we have
\begin{equation}
    u(x)+v(T(x))=\langle x, T(x) \rangle. 
    \label{eqn:maximizer_condition}
\end{equation}
This shows the existence of the solution.

From the definition of c-transform in Eqn.~(\ref{eqn:c_transform}), we obtain $v(y)=\inf_{x\in\bar{\Omega}} \langle x, y \rangle - u(x)$. Since $u(x)$ is concave and almost everywhere differentiable, we have $\nabla_x \langle x, y \rangle - \nabla u(x) = 0$, which implies that $y=T(x)=\nabla u(x)$. Therefore, the WT map is the gradient of the worst Brenier potential $u$. 

Next, we show the uniqueness of the WT map. Suppose there are two maximizers $(\varphi,\psi)\in K$ and $(u,v)\in K$, because $J(u,v)$ is linear,therefore $\frac{1}{2}(\varphi+u,\psi+v)\in K$ is also a maximizer. Assume 
\[
\begin{split}
\varphi(x_0)+\psi(y_0)=\langle x_0,y_0\rangle&, \varphi(x_0)+\psi(y) < \langle x_0, y \rangle, \forall y\neq y_0\\
u(x_0)+v(z_0)=\langle x_0,z_0\rangle&, u(x_0)+v(z) < \langle x_0, z \rangle, \forall z\neq z_0.
\end{split}
\]
If $y_0\neq z_0$, then  $\forall y$, $1/2(\varphi+u)(x_0)+1/2(\psi+v)(y) < \langle x_0,y \rangle$. But $(\frac{1}{2}(\varphi+u),\frac{1}{2}(\psi+v))$ is also a maximizer, this contradicts to the Eqn.~(\ref{eqn:maximizer_condition}). This shows the uniqueness of the WT map.

Finally, $u$ is concave and piecewise linear, therefore by Alexandrov's theorem \cite{Alexandrov2005Convex}, it is almost everywhere $C^2$. Moreover, the WT map $T=\nabla u$ is measure-preserving and $T_\#\mu=\nu$, thus we have
\[
\text{det}(DT)(x) = \frac{f(x)}{g\circ T(x)}\implies \text{det}(D^2 u)(x) = \frac{f(x)}{g\circ \nabla u(x)}
\]
This completes the proof.  $\square$
\end{proof}


\vspace{-4mm}
\subsection{Geometric Variational Method} 
\vspace{-1mm}
If $u$ is not smooth, we can still define the Alexandrov solution. The sub-gradient of a convex function $u$ at $x$ is defined as
\[
\partial u(x):= \left\{p\in\mathbb{R}^{d}: u(z) \ge \langle p, z-x \rangle + u(x), \forall z\in \Omega\right\}
\]
The sub-gradient defines a set-valued map: $\partial u:\Omega\to 2^{\Omega^*}$,  $x\mapsto \partial u(x)$. We can use the sub-gradient to replace the gradient map in Eqn.~(\ref{eqn:Monge_ampere}), and define 
\begin{definition}[Alexandrov Solution]A convex function $u:\Omega\to\mathbb{R}$ satisfies the equation $(\partial u)_\# \mu = \nu$, or $\mu( (\partial u)^{-1}(B)) = \nu(B)$, $\forall Borel~ set ~B\subset \Omega^*$, then $u$ is called an Alexandrov solution to the Monge-Amp\`ere equation Eqn.~(\ref{eqn:Monge_ampere}).
\end{definition}
The work of \cite{gu2013theory} proves a geometric variational approach for computing the Alexandrov solution of the optimal transportation problem.\\

\begin{figure}[t]
    \centering
    \begin{tabular}{ccccc}
    \includegraphics[height=0.2\textwidth]{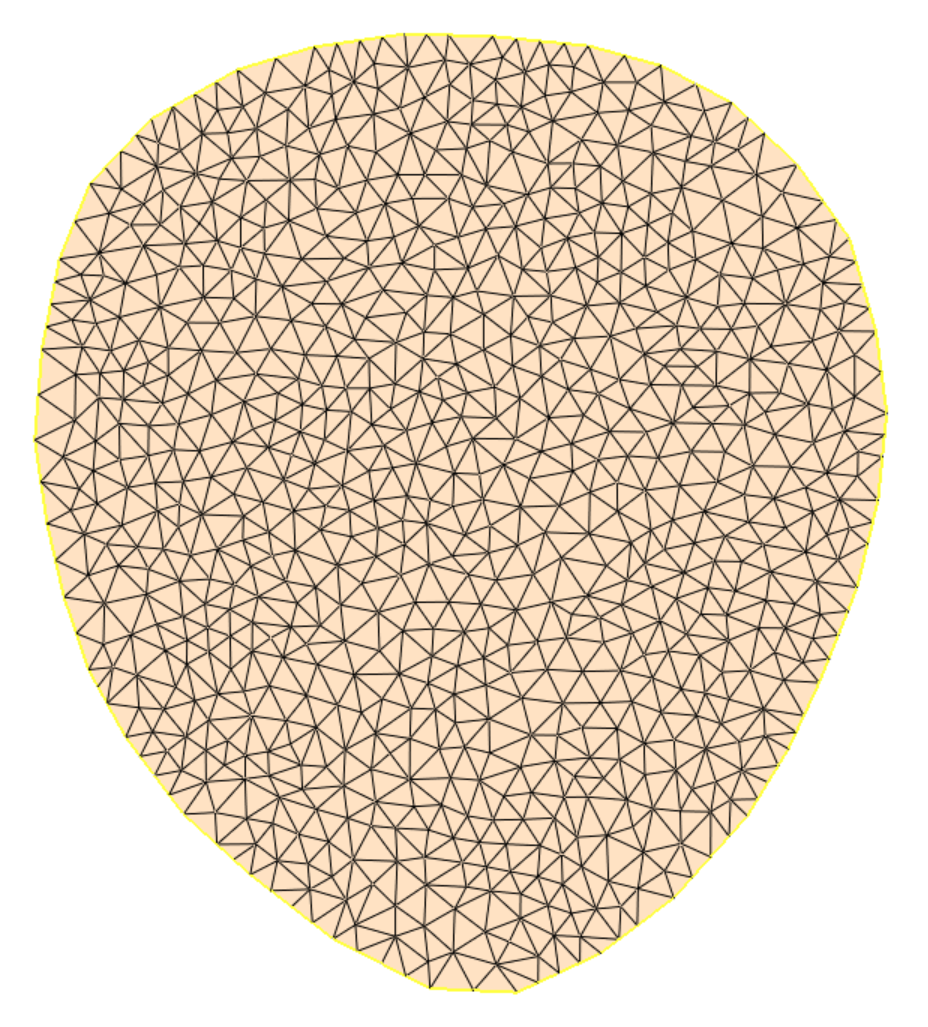}&
    \includegraphics[height=0.2\textwidth]{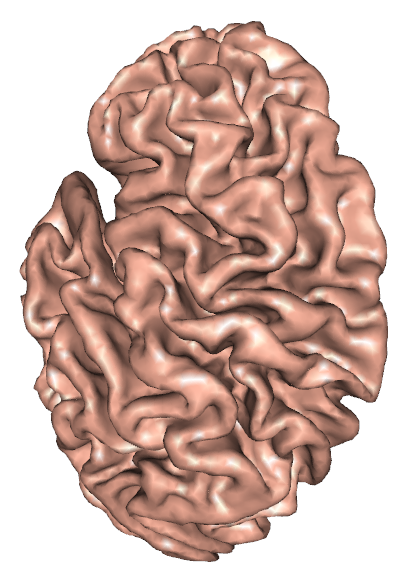}&
    \includegraphics[height=0.2\textwidth]{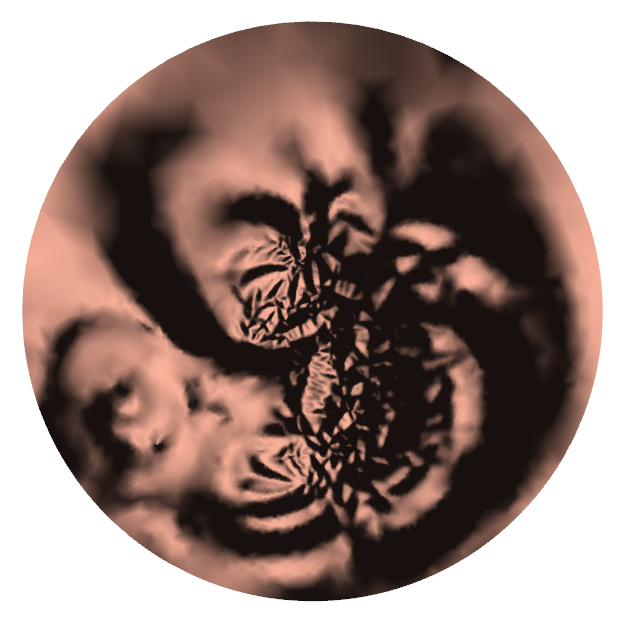} &
    \includegraphics[height=0.2\textwidth]{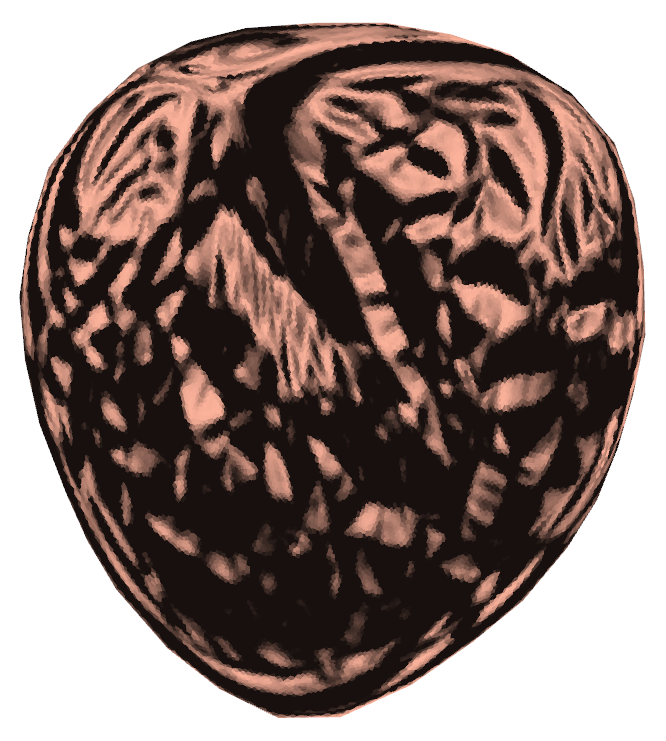}&
    \includegraphics[height=0.2\textwidth]{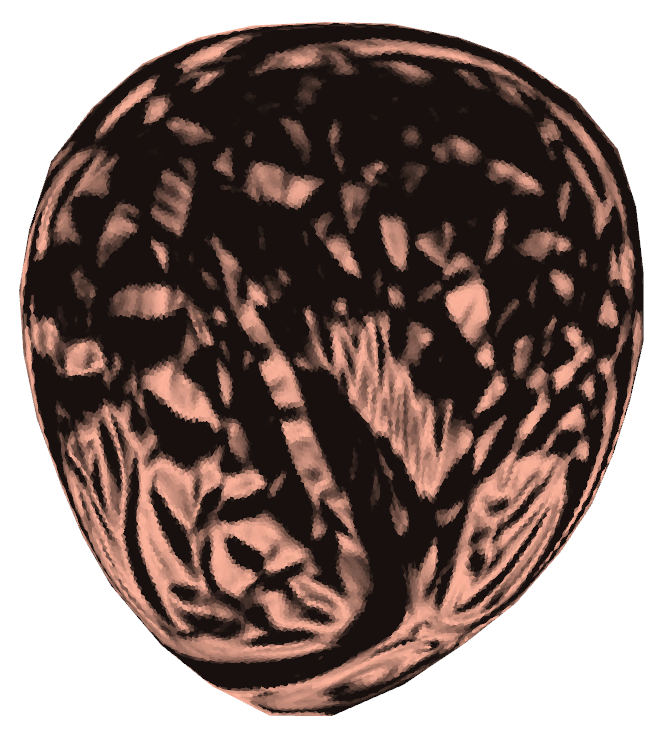}\\
    (a) $(\Omega,\mu)$ &(b) 3D surface &(c) $(\Omega^*,\nu)$ &(d) OT map image &(e) WT map image
    \end{tabular}
    \caption{The OT map and WT map from the source measure to the target measure given by the 3D surface.
    \label{fig:alex_input_measures}}
    \vspace{-4mm}
\end{figure}

\noindent{\bf Semi-discrete OT/WT maps}
Suppose the source measure is $(\Omega,\mu)$, $\Omega$ is a compact convex domain with non-empty interior in $\mathbb{R}^{d}$ and the density function $f(x)$ is continuous (Fig. \ref{fig:alex_input_measures}(a) gives an example). The target discrete measure $(\Omega^*,\nu)$ is defined as $\nu=\sum_{i=1}^n \nu_i \delta(y-p_i)$, where $p_i \subset \mathbb{R}^{d}$ are distinct $n$ points with $\nu_i >0$ and $\sum_{i=1}^n \nu_i = \mu(\Omega)$ (Fig. \ref{fig:alex_input_measures}(c) shows an example with the discrete measure coming from the 3D surface of Fig. \ref{fig:alex_input_measures}(b)). Alexandrov~\cite{Alexandrov2005Convex} claims that there exists a \emph{height vector} $\mathbf{h}=(h_1,\dots, h_n) \in \mathbb{R}^n$, so that the upper envelope $u_\mathbf{h}$ of the hyper-planes $\{\pi_i(x):=\langle x,p_i\rangle + h_i\}_{i=1}^n$ gives an open convex polytope $P(\mathbf{h})$, the volume of the projection of the $i$-th facet of $P(\mathbf{h})$ in $\Omega$ equals to $\nu_i$ $\forall i=1,2,\dots,n$. Furthermore, this convex polytope is unique up to a vertical translation. In fact, Yau's work \cite{gu2013theory} pointed out that the Alexandrov convex polytope $P(\mathbf{h})$, or equivalently the upper envelop $u_{\mathbf{h}}$ is exactly the Brenier potential, whose gradient gives the OT map shown in Fig. \ref{fig:alex_input_measures}(d).

\begin{theorem}[Yau et. al. \cite{gu2013theory}]  Let $\Omega \in \mathbb{R}^{d}$ be a compact convex domain, $\{p_1, ..., p_n\}$ be a set of distinct points in $\mathbb{R}^d$ and $f: \Omega \to \mathbb{R}$ be a positive continuous function. Then for any $\nu_1,\dots, \nu_n >0$ with $\sum_{i=1}^n \nu_i =\int_{\Omega} f(x) dx$, there exists $\mathbf{h}=(h_1, h_2,\dots, h_n) \in \mathbb{R}^n$, unique up to adding a constant $(c,c,\dots, c)$, so that $\mu(W_i(\mathbf{h})\cap\Omega) = \int_{W_i(\mathbf{h}) \cap \Omega}f(x) dx  =\nu_i$, $\forall~ i=1,2,\ldots,n$. The height vector $\mathbf{h}$ is exactly the minimum of the following convex function
\begin{equation}
E(\mathbf{h}) = \int^h_0 \sum_{i=1}^n \mu(W_i(h)\cap\Omega)  dh_i -\sum_{i=1}^n h_i \nu_i
\label{eqn:energyE}
\end{equation}
on the open convex set (admissible solution space)
\begin{equation}
    \mathcal{H} =\{ \mathbf{h} \in \mathbb{R}^n | \mu(W_i(\mathbf{h}) \cap \Omega) >0~ \forall i=1,2,\ldots,n\}\bigcap \left\{\mathbf{h} \in \mathbb{R}^n |\sum_{i=1}^n h_i = 0\right\}.
    \label{eqn:admissible_space}
\end{equation}
Furthermore, the gradient map $\nabla u_\mathbf{h}$ minimizes the quadratic cost 
$\frac{1}{2} \int_{\Omega} | x - T(x)|^2 f(x) dx$ among all the measure preserving maps $T: (\Omega, \mu) \to (\mathbb{R}^d$, $\nu=\sum_{i=1}^n \nu_i \delta_{p_i})$, $T_\#\mu=\nu$.
\label{thm:Yau}
\end{theorem}

\begin{theorem}[Semi-Discrete Worst Transportation Map]  Let $\Omega\in\mathbb{R}^{d}$ be a compact convex domain, $\{p_1, ..., p_n\}$ be a set of distinct points in $\mathbb{R}^{d}$ and $f: \Omega \to \mathbb{R}$ be a positive continuous function. Then for any $\nu_1,\dots, \nu_n >0$ with $\sum_{i=1}^n \nu_i =\int_{\Omega} f(x) dx$, there exists $\mathbf{h}=(h_1, h_2,\dots, h_n) \in \mathbb{R}^n$, unique up to adding a constant $(c,c,\dots, c)$, so that $\mu(W_i(\mathbf{h})\cap\Omega) = \int_{W_i(\mathbf{h}) \cap \Omega}f(x) dx  =\nu_i$, $\forall i$. The height vector $\mathbf{h}$ is exactly the {\bf maximum} of the following concave function
\begin{equation}
E(\mathbf{h}) = \int^h_0 \sum_{i=1}^n \mu(W_i(\mathbf{h})\cap\Omega)  dh_i -\sum_{i=1}^n h_i \nu_i
\label{eqn:energyE}
\end{equation}
on the open convex set (admissible solution space) defined on Eqn. (\ref{eqn:admissible_space}).
Furthermore, the gradient map $\nabla u_{\mathbf{h}}$ {\bf maximizes} the quadratic cost $\frac{1}{2} \int_{\Omega} | x - T(x)|^2 f(x) dx$ among all the measure preserving maps $T: (\Omega, \mu) \to (\mathbb{R}^{d}$, $\nu=\sum_{i=1}^n \nu_i \delta_{p_i})$, $T_\#\mu=\nu$.
\label{thm:wtp}
\end{theorem}
\begin{proof} Given the height vector $\mathbf{h}=(h_1,h_2,\cdots,h_n)$, $\mathbf{h}\in \mathcal{H}$, we construct the upper convex hull of $v_i(\mathbf{h})=(p_i,-h_i)$'s (see Fig.~\ref{fig:OT_WT_Legendre_duals}(d)), each vertex corresponds to a plane $\pi_i(\mathbf{h},x):= \langle p_i, x\rangle + h_i$. The convex hull is dual to the lower envelope of the plane $\pi_i(\mathbf{h},\cdot)$ (see Fig.~\ref{fig:OT_WT_Legendre_duals}(c)), which is the graph of the concave function $u_\mathbf{h}(x) := \min_{i=1}^n \left\{ \langle p_i, x\rangle + h_i \right\}$.
The projection of the lower envelope induces a farthest power diagram $\mathcal{D}(\mathbf{h})$ (see Fig.~\ref{fig:OT_WT_Legendre_duals}(c)) with $\Omega = \bigcup_{i=1}^n W_i(\mathbf{h})\cap \Omega$, $W_i(\mathbf{h}) := \{x\in \mathbb{R}^{d}$ and $\nabla u_{\mathbf{h}}(x)=p_i\}$. 

\begin{figure}[t]
    \centering
    \setlength{\tabcolsep}{2pt}
    \begin{tabular}{cccc}
    \includegraphics[height=0.15\textwidth]{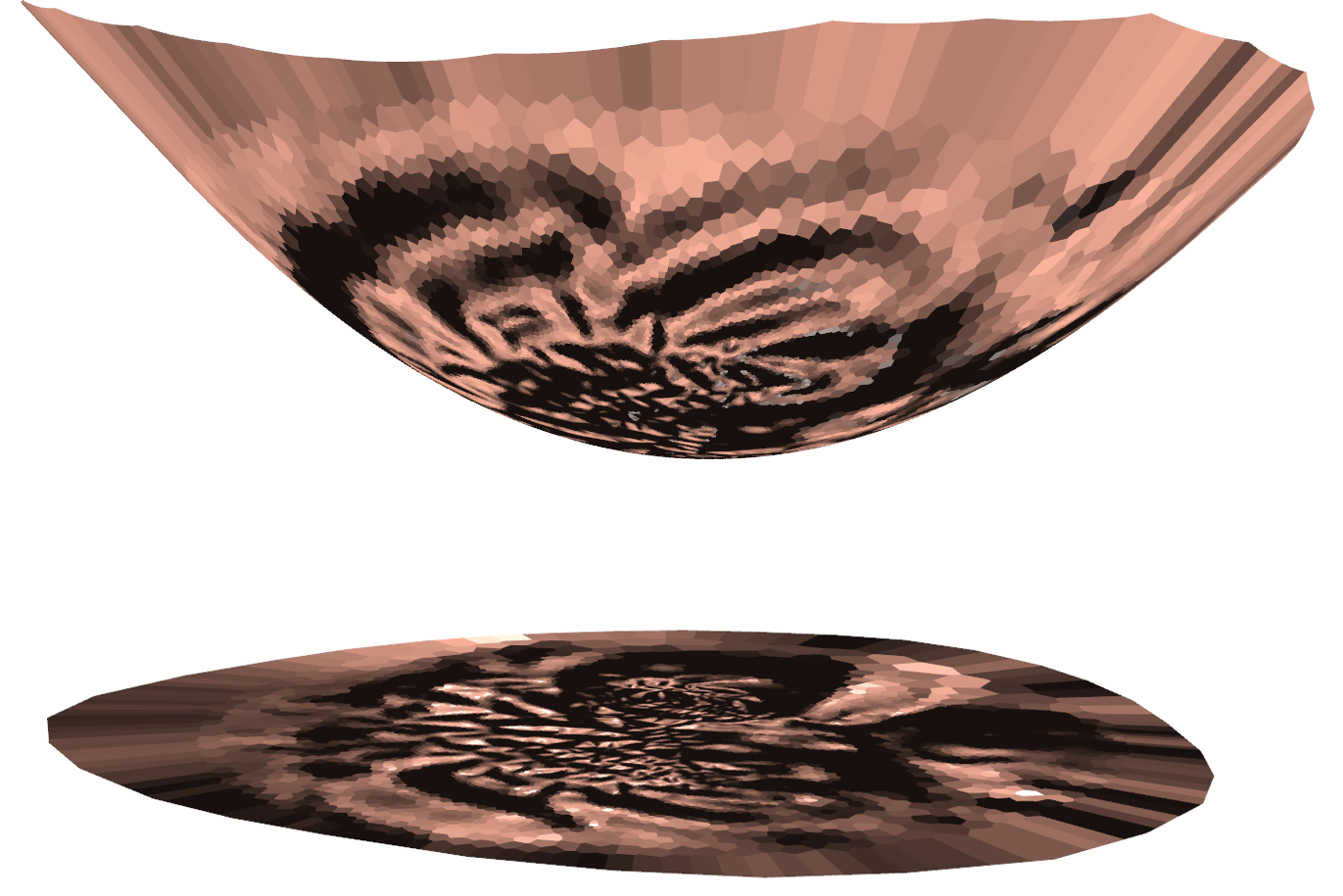}&
    \includegraphics[height=0.15\textwidth]{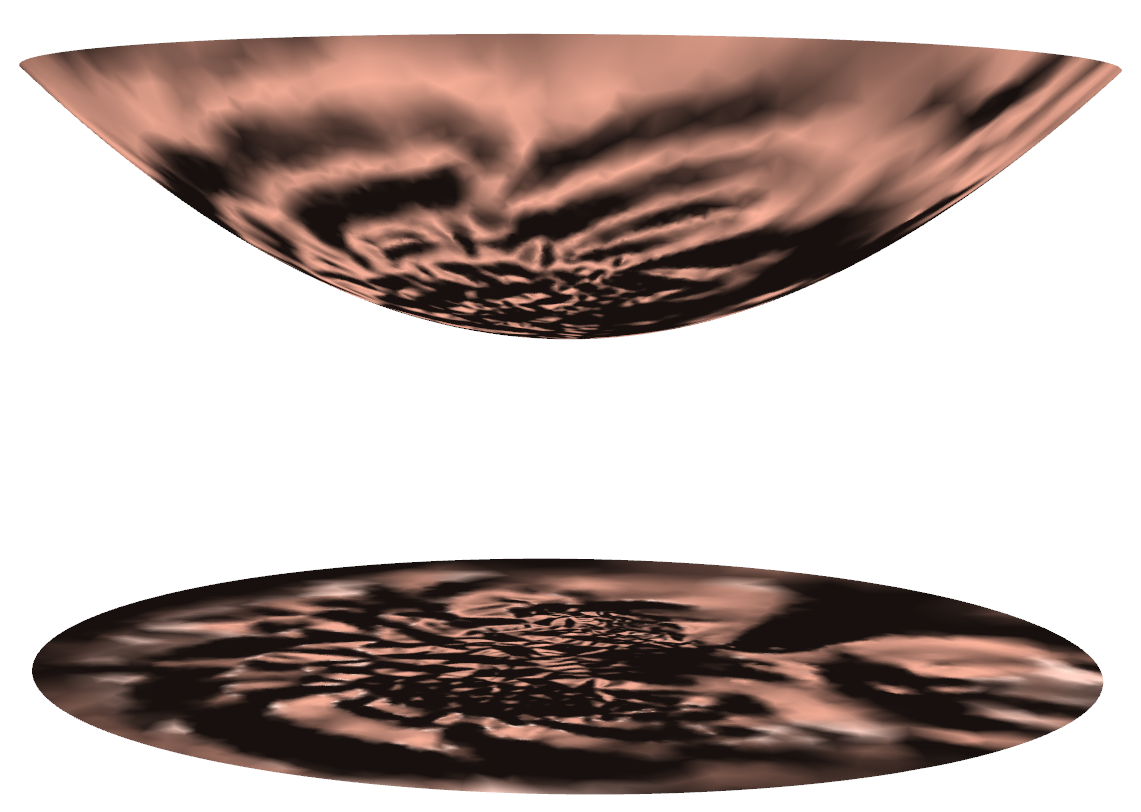}& 
    \includegraphics[height=0.15\textwidth]{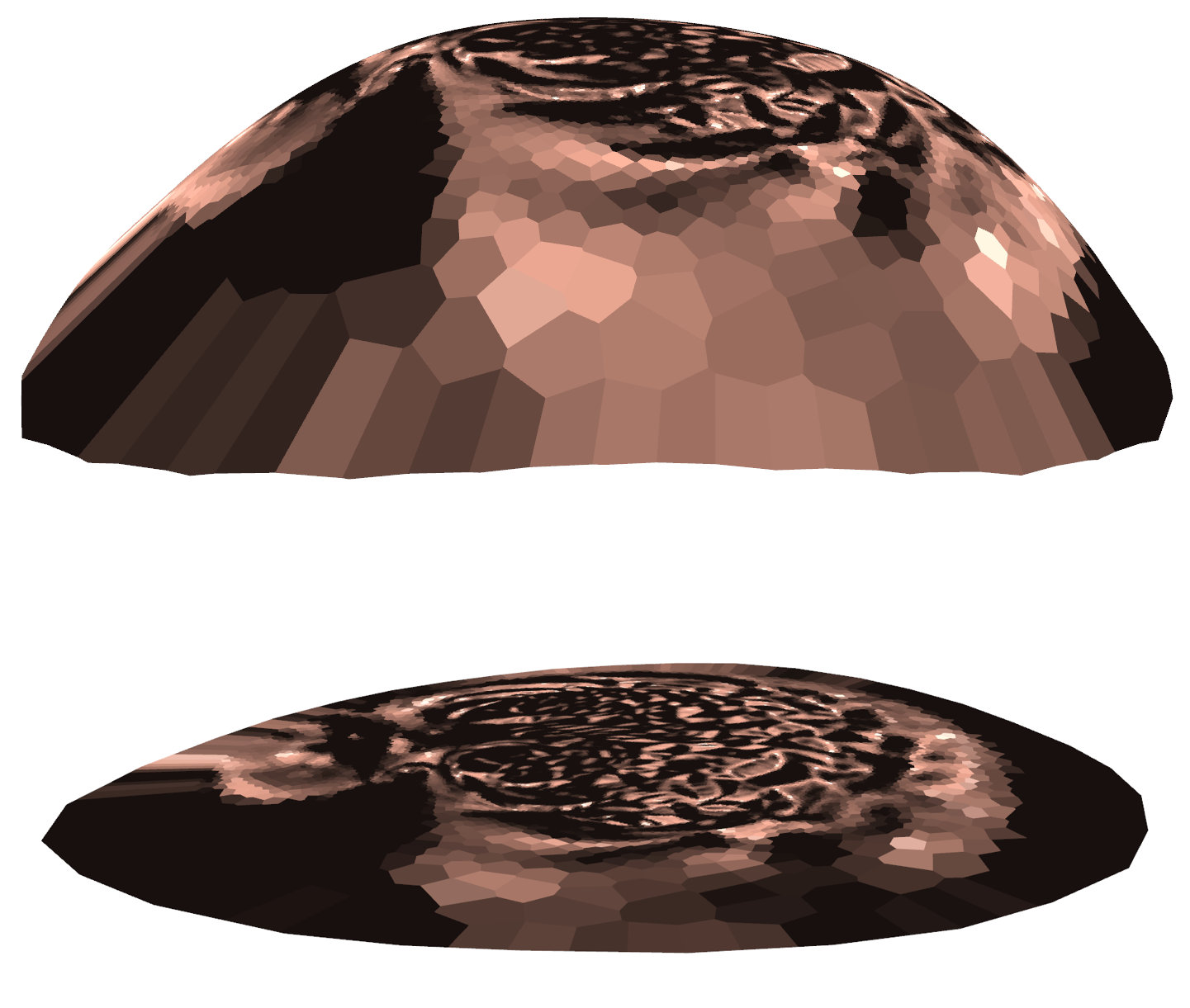}&
    \includegraphics[height=0.15\textwidth]{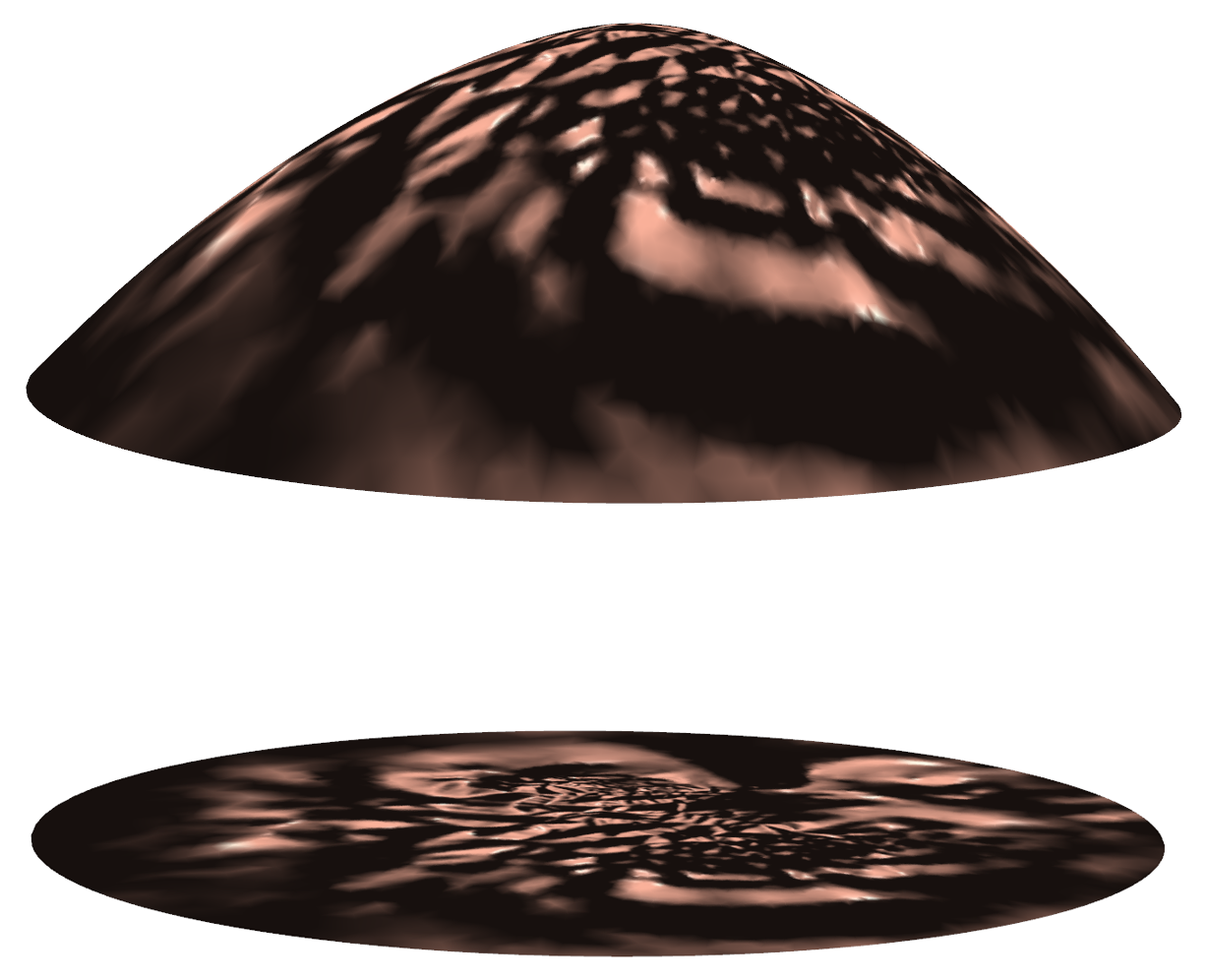}\\
    (a) OT envelope &(b) OT convex hull &
    (c) WT envelope &(d) WT concave hull\\
    \end{tabular}
    \caption{
    The Brenier potential for OT map and WT map, equivalently the upper (a) and lower (c) envelopes. The Legendre dual of the potential of the OT map and WT map, equivalently the lower (b) and upper (d) convex hulls.
    \label{fig:OT_WT_Legendre_duals}}
    \vspace{-5mm}
\end{figure}

The $\mu$-volume of each cell is defined as 
\begin{equation}
    w_i(\mathbf{h}) := \mu(W_i(\mathbf{h})\cap \Omega) = \int_{W_i(\mathbf{h})\cap\Omega} f(x) dx.
    \label{eqn:mu_cell_volume}
\end{equation}
Similar to Lemma 2.5 in \cite{gu2013theory}, by direct computation we can show the symmetric relation holds:
\begin{equation}
    \frac{\partial w_i(\mathbf{h})}{\partial h_j} = \frac{\partial w_j(\mathbf{h})}{\partial h_i} = \frac{1}{|p_i-p_j|} \int_{W_i(\mathbf{h})\cap W_j(\mathbf{h})\cap\Omega} f(x) ds.
    \label{eqn:mu_edge_length}
\end{equation}
This shows the differential form $\omega = \sum_{i=1}^n w_i(\mathbf{h})dh_i$ is a closed one-form. As in ~\cite{gu2013theory}, by Brunn-Minkowski inequality, one can show that the admissible height space $\mathcal{H}$ in Eqn.~(\ref{eqn:admissible_space}) is convex and simply connected. Hence $\omega$ is exact. So the energy 
$E(\mathbf{h}):= \int_{0}^{\mathbf{h}} \omega$ is well defined and its Hessian matrix is given by 
\begin{equation}
    \frac{\partial^2 E(\mathbf{h})}{\partial h_i\partial h_j} = \frac{w_i(\mathbf{h})}{\partial h_j} \ge 0,
    \label{eqn:Hessian_1}
\end{equation}
Since the total volume of all the cells is the constant $\mu(\Omega)$, we obtain
\begin{equation}
    \frac{\partial^2 E(\mathbf{h})}{\partial h_i^2} = -\sum_{j\neq i} \frac{w_i(\mathbf{h})}{\partial h_j}<0.
    \label{eqn:Hessian_2}
\end{equation}
Therefore, the Hessian matrix is negative definite in $\mathcal{H}$ and the energy is strictly concave in $\mathcal{H}$. By adding a linear term, the following energy is still strictly concave,
\[
    E(\mathbf{h}) = \int_{0}^{\mathbf{h}} \sum_{i=1}^n w_i(\mathbf{h}) dh_i - \sum_{i=1}^n \nu_i h_i.
\]
The gradient of $E(\mathbf{h})$ is given by
\begin{equation}
    \nabla E(\mathbf{h}) = (w_1(\mathbf{h})  - \nu_1, w_2(\mathbf{h})  - \nu_2, \cdots, w_n(\mathbf{h})  - \nu_n).
    \label{eqn:gradient}
\end{equation}
On the boundary of $\mathcal{H}$, there is an empty cell $W_k(\mathbf{h})$, and the $k$-th component of the gradient  is $-\nu_k$, which points to the interior of $\mathcal{H}$. This shows that the global unique maximum of the energy is in the interior of $\mathcal{H}$. At the maximum point $\mathbf{h}^*$, $\nabla E(\mathbf{h}^*)$ is zero and $w_i(\mathbf{h}^*)=\nu_i$. Thus, $\mathbf{h}^*$ is the unique solution to the semi-discrete worst transportation problem, as shown in Fig. \ref{fig:alex_input_measures}(e) $\square$
\end{proof}

\vspace{-5mm}
\section{Computational Algorithms}
\label{sec:algorithm}
\vspace{-3mm}

This section gives a unified algorithm to compute both the optimal and the worst transportation maps based on convex geometry \cite{Berg2008CG}.
\vspace{-4mm}
\subsection{Basic Concepts from Computational Geometry}
\vspace{-1mm}
A hyperplane in $\mathbb{R}^{d+1}$ is represented as $\pi(x):=\langle p,x \rangle + h$. Given a family of hyperplanes $\{\pi_i(x)=\langle p_i,x\rangle + h_i\}_{i=1}^n$, their \emph{upper envelope} of $\{\pi_i\}_{i=1}^n$ is the graph of the function 
$u(x) := \max_{i=1}^n \left\{\langle p_i,x \rangle + h_i\right\}$; 
the \emph{lower envelope} is the graph of the function
$u(x) := \min_{i=1}^n \left\{\langle p_i,x \rangle + h_i\right\}$;
the Legendre dual of $u$ is defined as $u^*(y):= \max_{x\in \mathbb{R}^d} \langle x, y \rangle - u(x)$. The c-transform of $u$ is defined as
\begin{equation}
    u^c(y):= \min_{x\in \mathbb{R}^{d}} \langle x, y \rangle - u(x). 
\end{equation}
Each hyperplane $\pi_i(x)$ has a dual point in $\mathbb{R}^{d+1}$, namely $\pi_i^*:=(p_i,-h_i)$. The graph of $u^*$ is the \emph{lower convex hull} of the dual points $\{\pi_i^*\}_{i=1}^n$. And the graph of $u^c$ is the \emph{upper convex hull} of the dual points $\{\pi_i^*\}_{i=1}^n$. (\textit{i}) The projection of the upper envelope induces a \emph{nearest power diagram} $\mathcal{D}(\Omega)$ of $\Omega$ with $\Omega=\bigcup_{i=1}^n W_i(\mathbf{h})$ and $W_i(\mathbf{h}):=\left\{x\in \Omega| \nabla u(x)=p_i\right\}$. And the projection of the lower convex hull $u^*$ induces a \emph{nearest weighted Delaunay triangulation} $\mathcal{T}(\Omega^*)$ of $\Omega^*$.
(\textit{ii}) The projection of the lower envelope induces a \emph{farthest power diagram} $\mathcal{D}^c$ of $\Omega$. And the projection of the upper convex hull $u^c$ induces a \emph{farthest weighted Delaunay triangulation} $\mathcal{T}^c(\Omega^*)$. $\mathcal{D}(\Omega)$ and $\mathcal{T}(\Omega^*)$ are dual to each other, namely $p_i$ connects $p_j$ in $\mathcal{T}(\Omega^*)$ if and only if $W_i(\mathbf{h})$ is adjacent to $W_j(\mathbf{h})$. Similarly, $\mathcal{D}^c$ and $\mathcal{T}^c$ are also dual to each other. Fig.~\ref{fig:OT_WT_Legendre_duals} shows these basic concepts.\\
\vspace{-4mm}
\subsection{Algorithms based on Computational Geometry}
\vspace{-1mm}
\paragraph{Pipeline}
The algorithm in Alg. \ref{alg:WT-OT} mainly optimizes the energy $E(\mathbf{h})$ in the admissible solution space $\mathcal{H}$ using Newton's method. At the beginning, for the OT (WT) map, the height vector $\mathbf{h}_0$ is initialized as $h_i=-\frac{1}{2}|p_i|^2$ ($h_i=\frac{1}{2}|p_i|^2$). At each step, the convex hull of $\{(p_i, -h_i)\}_{i=1}^n$ is constructed. For the OT (WT) map, the lower (upper) convex hull is projected to induce a nearest (farthest) weighted Delaunay triangulation $\mathcal{T}$ of $\{p_i\}$'s. Each vertex $v_i(\mathbf{h})=(p_i,-h_i)$ on the convex hull  corresponds to a supporting plane $\pi_i(\mathbf{h},x)=\langle p_i, x\rangle + h_i$, each face $[v_i,v_j,v_k]$ in the convex hull is dual to the vertex in the envelope, which is the intersection point of $\pi_i,\pi_j$ and $\pi_k$. For the OT (WT) map, the lower (upper) convex hull is dual to the upper (lower) envelope, and the upper (lower) envelope induces the nearest (farthest) power diagram. The relationship of the convex hulls and the envelopes are shown in Fig.~\ref{fig:OT_WT_Legendre_duals}.

\begin{algorithm}[!ht]
\SetAlgoLined
\KwIn{$(\Omega,\mu)$, $\{(p_i,\nu_i)\}_{i=1}^n$}
\KwOut{The optimizer $\mathbf{h}$ of the Brenier potential $u_{\mathbf{h}}$}
 Normalize $\{p_1,p_2,\dots,p_n\}$ to be inside $\Omega$ by translation and scaling\;
 Initialize $h_i = \pm \langle p_i, p_i \rangle /2$ for WT/OT\;
 \While{true}{
    Compute the upper (lower) convex hull of $\{(p_i,-h_i)\}_{i=1}^n$ for WT/OT map\;
    Compute the lower (upper) envelope of the planes $\{\langle p_i,x \rangle + h_i\}_{i=1}^n$ for WT/OT map\;
    Project the lower (upper) envelope to $\Omega$ to get the farthest (nearest) power diagram $\Omega = \bigcup_{i=1}^n W_i(\mathbf{h})$ for WT/OT map \;
    Compute the $\mu$-volume of each cell $w_i(\mathbf{h})= \mu(W_i(\mathbf{h}))$ using Eqn.~(\ref{eqn:mu_cell_volume})\;
    Compute the gradient of the energy $E(\mathbf{h})$, $\nabla E(\mathbf{h})=(w_i(\mathbf{h})-\nu_i)$\;
  \If{$\|\nabla E(\mathbf{h})\|<\varepsilon$}{
        \Return $\mathbf{h}$\;
   }
   Compute the $\mu$-lengths of the power Voronoi edges $W_i(\mathbf{h})\cap W_j(\mathbf{h})\cap\Omega$ using Eqn.~(\ref{eqn:mu_edge_length})\;
   Construct the Hessian matrix of the energy $E(\mathbf{h})$ for WT/OT map:
   \[
        \text{Hess}(E(\mathbf{h})):=\frac{\partial^2 E(\mathbf{h})}{\partial h_i\partial h_j} = \pm\frac{\mu(W_i(\mathbf{h})\cap W_j(\mathbf{h}))}{|y_i-y_j|}
   \]
   Solve the linear system: $\text{Hess}(E(\mathbf{h})) \mathbf{d} = \nabla E(\mathbf{h})$\;
   $\lambda\leftarrow \pm 1$ for WT/OT map\;
   \Repeat{no empty power cell}{
   Compute the farthest (nearest) power diagram $\mathcal{D}(\mathbf{h}+\lambda\mathbf{d})$ for WT/OT map\;
    $\lambda\leftarrow \frac{1}{2}\lambda$\;
   }
   Update the height vector
   $\mathbf{h} \leftarrow \mathbf{h} + \lambda \mathbf{d}$\;
 }
\caption{Worst/Optimal Transportation Map}
\label{alg:WT-OT}
\end{algorithm}

Then we compute the $\mu$-volume of each power cell using Eqn.~(\ref{eqn:mu_cell_volume}), the gradient of the energy Eqn.~(\ref{eqn:energyE}) is given by Eqn.~(\ref{eqn:gradient}). The Hessian matrix $\text{Hess}(E(\mathbf{h}))$ can be constructed using Eqn.~(\ref{eqn:Hessian_1}) for off diagonal elements and Eqn.~(\ref{eqn:Hessian_2}) for diagonal elements. The Hessian matrices of the OT map and the WT map differ by a sign. 
Then we solve the following linear system to find the update direction,
\begin{equation}
    \text{Hess}(E(\mathbf{h})) \mathbf{d} = \nabla E(\mathbf{h}).
\end{equation}
Next we need to determine the step length $\lambda$, such that $\mathbf{h}+\lambda\mathbf{d}$ is still in the admissible solution space $\mathcal{H}$ in Eqn. (\ref{eqn:admissible_space}). Firstly, we set $\lambda=-1$ for OT map and $\lambda=+1$ for WT map. Then we compute the power diagram $\mathcal{D}(\mathbf{h}+\lambda \mathbf{d})$. If some cells disappear in $\mathcal{D}(\mathbf{h}+\lambda \mathbf{d})$, then it means $\mathbf{h}+\lambda\mathbf{d}$ exceeds the admissible space. In this case, we shrink $\lambda$ by half, $\lambda\leftarrow \frac{1}{2}\lambda$, and recompute the power diagram with $\mathbf{h}+\lambda\mathbf{d}$. We repeat this process to find an appropriate step length $\lambda$ and update $\mathbf{h}=\mathbf{h}+\lambda\mathbf{d}$. We repeat the above procedures until the norm of the gradient $\|\nabla E(\mathbf{h})\|$ is less than a prescribed threshold $\varepsilon$. As a result, the upper (lower) envelope is the Brenier potential, the desired OT(WT) mapping maps each nearest (farthest) power cell $W_i(\mathbf{h})$ to the corresponding point $p_i$.

\paragraph{Convex Hull}
In order to compute the power diagram, we need to compute the convex hull \cite{Berg2008CG}. The conventional method \cite{Su2015Registration} computes the convex hull from the scratch at each iteration, which is the most time-consuming step in the algorithm pipeline. Actually, at the later stages of the optimization, the combinatorial structure of the convex hull does not change much. Therefore, in the proposed method, we only locally update the connectivity of the convex hull. Basically, we check the local power Delaunay property of each edge, and push the non-Delaunay edges to a stack. While the stack is non-empty, we pop the top edge and check whether it is local power Delaunay, if it is then we continue, otherwise we flip it. Furthermore, if the flipping causes some overlapped triangles, we flip it back. By repeating this procedure, we will finally update the convex hull and project it to the weighted Delaunay triangulation. If in the end, the stack is empty, but there are still non-local power Delaunay edges, then it means that the height vector $\mathbf{h}$ is outside the admissible space $\mathcal{H}$ and some power cells are empty. In this scenario, we reduce the step length $\lambda$ by half and try again.

\begin{figure}[t]
\vspace{-2mm}
    \centering
    \begin{tabular}{ccc}
    \includegraphics[width=0.25\textwidth]{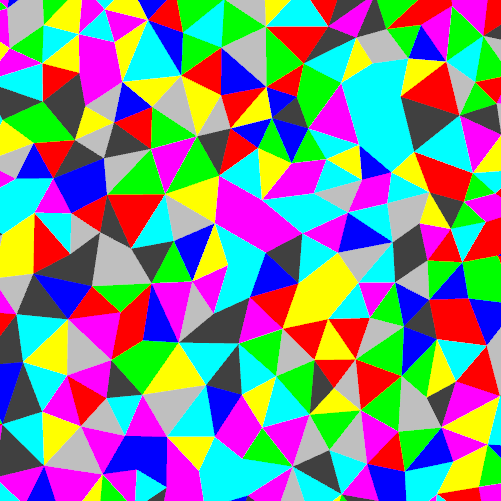}&
    \includegraphics[width=0.25\textwidth]{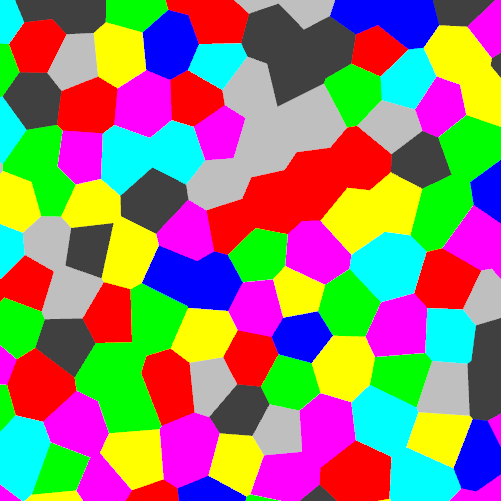}&
    \includegraphics[width=0.25\textwidth]{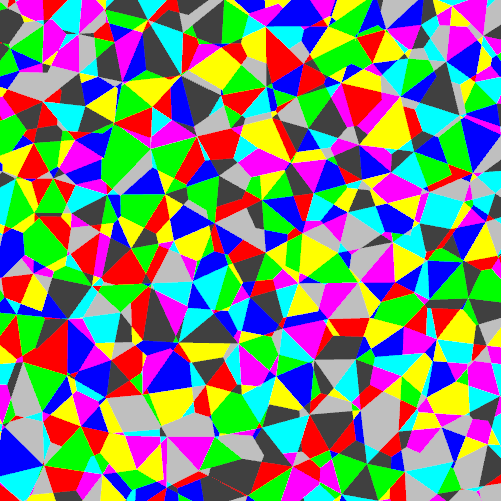}\\
    (a) PL source density  &(b) Power diagram &(c) Overlay \\
    \end{tabular}
        \caption{The subdivision algorithm computes the subdivision of the source triangulation, where the density function is defined, and the power diagram.}
        \label{fig:subdivision}
        \vspace{-4mm}
\end{figure}

\paragraph{Subdivision}
With a piecewise linear source density, we need to compute the $\mu$-area of the power cells and the $\mu$-length of the power diagram edges. The source measure is represented by a piecewise linear density function, defined on a triangulation, as shown in Fig. \ref{fig:subdivision}(a). Therefore, we need to compute the overlay (Fig. \ref{fig:subdivision}(c)) of the triangulation (Fig. \ref{fig:subdivision}(a)) and the power diagram (Fig. \ref{fig:subdivision}(b)) in each iteration. This step is technically challenging. If we use a naive approach to compute the intersection between each triangle and each power cell, the complexity is very high. Therefore, we use a Bentley-Ottmann type sweep line approach \cite{Bentley-Ottmann1979} to improve the efficiency. Basically, all the planar points are sorted, such that the left-lower points are less than the right-upper ones. Then for each cell we find the minimal vertex and maximal vertex. A sweep line goes from left to right. When the sweep line hits the minimal vertex of a cell, the cell is born; when the sweep line passes the maximal vertex, the cell dies. We also keep the data structure to store all the alive triangles and cells, and compute the intersections among them. This greatly improves the computation efficiency.

\vspace{-5mm}
\section{Experiments}
\vspace{-3mm}
To show the practicality of our framework for structural MR images as well as the robustness over large brain image datasets, we aim to use the WT cost to statistically discriminate A$\beta+$ AD patients and A$\beta-$ CU subjects.

\textbf{Data Preparation} Brain sMRI data are obtained from the Alzheimer’s Disease Neuroimaging Initiative (ADNI) database \cite{ANDI}, from which we use 81 A$\beta+$ AD patients and 110 A$\beta-$ CU subjects. The ADNI florbetapir PET data is processed using AVID pipeline~\cite{Navitsky:AD18} and later converted to centiloid scales. A centiloid cutoff of 37.1 is used to determine amyloid positivity~\cite{Fleisher:an2011}.  The sMRIs are preprocessed using FreeSurfer~\cite{freesurfer}
to reconstruct the pial cortical surfaces and we only use the left cerebral surfaces.
For each surface, we remove the corpus callosum region which has little morphometry information related to AD, so that the final surface becomes a topological disk.
Further, we compute the conformal mapping $\phi$ from the surface $S$ to the planar disk $\mathbb{D}$ with the discrete surface Ricci flow \cite{WangRiccTMI}. To eliminate the Mobius ambiguity, we map the vertex with the largest $z$ coordinate on $S$ to the origin of the disk, and the vertex with the largest $z$ coordinate on the boundary $\partial S$ to $(0,1)$ coordinate of $\mathbb{D}$.

\textbf{Computation Process} In the experiment, we randomly select one subject from the A$\beta-$ CU subjects as the template, and then compute both the OT costs and WT costs from the template to all other cortical surfaces. The source measure is piecewisely defined on the parameter space of the template, namely the planar disk. The measure of each triangle $[\phi(v_i), \phi(v_j), \phi(v_k)]$ on the disk is equal to the corresponding area of the triangle $[v_i, v_j, v_k]$ on $S$. Then the total source measure is normalized to be $1$. For the target surface $M$, the measure is defined on the planar points $y_i=\phi(v_i)$ with $v_i\in M$. The discrete measure $\nu_i$ corresponding to both $v_i$ and $y_i$ is given by $\nu_i = \frac{1}{3}\sum_{[v_i, v_j, v_k]\in M} area([v_i, v_j, v_k])$, where $[v_i, v_j,v_k]$ represents a face adjacent to $v_i$ on $M$ in $R^3$. After normalization, the summation of the discrete measures will be equal to the measure of the planar disk, namely $1$. Then we compute both the OT cost and the WT cost from the planar source measure induced by the template surface to the target discrete measures given by the ADs and CUs. Finally, we run a permutation test with 50,000 random assignments of subjects to groups to estimate the statistical significance of both measurements. Furthermore, we also compute surface areas and cortical volumes as the measurements for comparison and the same permutation test is applied. The $p$-value for the WT cost is 2e-5, which indicates that the WT-based univariate biomarkers are statistically significant between two groups and they may be used as a reliable indicator to separate two groups. It is worth noting that it is also far better than that of the surface area, cortical volume and OT cost, which are given by $0.7859$, $0.5033$ and $0.7783$, respectively, as shown in Tab. \ref{tab:results}. 

\begin{table}[t]
\centering
\small
\caption{The permutation test results with surface area, surface volume, OT cost and WT cost for group difference between those of A$\beta+$ ADs and A$\beta-$ CUs.}
\begin{tabular}{c|c|c|c|c}
\hline
Method      & Surface Area &   Cortical Volume & OT Cost   & WT Cost  ~ \\ \hline
\textit{p-value}  &  0.7859  &  0.5033   &  0.7783   &  \textbf{2e-5}     \\ \hline
\end{tabular}
\label{tab:results}
\vspace{-6mm}
\end{table}

Our results show that the WT cost is promising as an AD imaging biomarker. It is unexpected to see that it performs far better than OT. A plausible explanation is that although human brain cortical shapes are generally homogeneous, the AD-induced atrophy may have some common patterns which are consistently exaggerated in the WT cost but OT is robust to these changes. More experiments and theoretical study are warranted to validate our observation.

\vspace{-5mm}
\section{Conclusion}
\vspace{-3mm}
In this work, we propose a new algorithm to compute the WT cost and validate its potential as a new AD imaging biomarker. In the future, we will validate our framework with more brain images and offer more theoretical interpretation to its improved statistical power.


\vspace{-5mm}
\bibliographystyle{abbrv}
\bibliography{ref}
\end{document}